\documentclass[journal=jacsat,manuscript=article]{achemso}

\usepackage[version=3]{mhchem} 
\usepackage{siunitx}
\usepackage{soul}

\author{Ofir Blumer}
\affiliation[TAU]
{School of Chemistry, Tel Aviv University, Tel Aviv 6997801, Israel.}
\author{Shlomi Reuveni}
\affiliation[TAU]
{School of Chemistry, Tel Aviv University, Tel Aviv 6997801, Israel.}
\alsoaffiliation[Sackler]
{The Center for Computational Molecular and Materials
Science, Tel Aviv University, Tel Aviv 6997801, Israel.}
\alsoaffiliation[BioSoft]
{The Center for Physics and Chemistry of Living Systems, Tel Aviv University, Tel Aviv 6997801, Israel.}
\author{Barak Hirshberg}
\email{hirshb@tauex.tau.ac.il}
\affiliation[TAU]
{School of Chemistry, Tel Aviv University, Tel Aviv 6997801, Israel.}
\alsoaffiliation[Sackler]
{The Center for Computational Molecular and Materials
Science, Tel Aviv University, Tel Aviv 6997801, Israel.}

\title{Stochastic Resetting for Enhanced Sampling}

\keywords{Enhanced sampling, Stochastic resetting, Molecular dynamics}

\begin{document}

\begin{abstract}
We present a method for enhanced sampling of molecular dynamics simulations using stochastic resetting. Various phenomena, ranging from crystal nucleation to protein folding, occur on timescales that are unreachable in standard simulations. This is often caused by broad transition time distributions in which extremely slow events have a non-negligible probability. Stochastic resetting, i.e., restarting simulations at random times, was recently shown to significantly expedite processes that follow such distributions. 
Here, we employ resetting for enhanced sampling of molecular simulations for the first time. We show that it accelerates long-timescale processes by up to an order of magnitude in examples ranging from simple models to molecular systems. Most importantly, we recover the mean transition time without resetting -- typically too long to be sampled directly -- from accelerated simulations at a single restart rate.
Stochastic resetting can be used as a standalone method or combined with other sampling algorithms to further accelerate simulations.
\end{abstract}

\section{Introduction}

\begin{figure}
  \includegraphics[width=\linewidth]{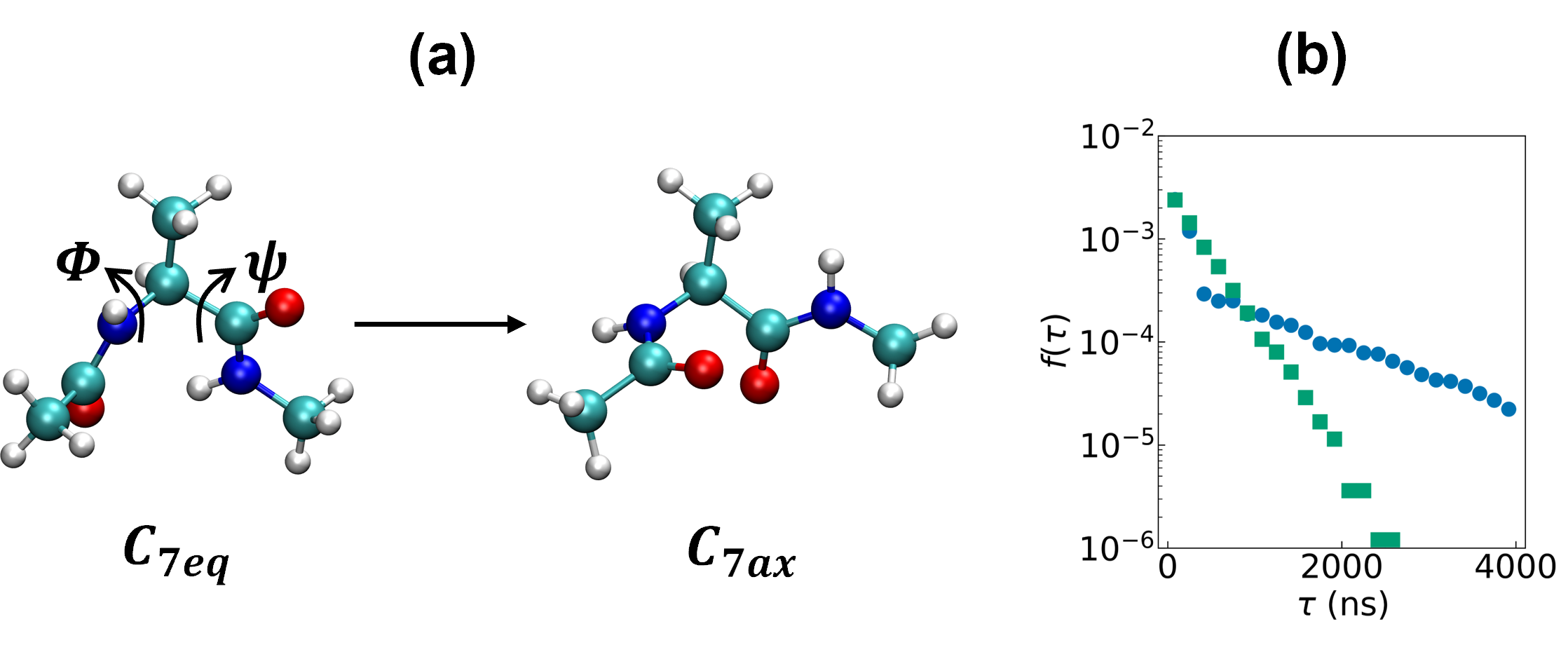}
  \caption{(a) The two conformers of an alanine dipeptide molecule. (b) The FPT distributions for transitions between them, starting from $C_{7eq}$, without resetting (blue circles) and with Poisson resetting at a rate of $r = 0.1 \, ns^{-1}$ (green squares).  The y axis is given on a logarithmic scale. 
  The full details of the simulation protocol and how the FPT was determined are given in the SI.}
  \label{fig:FPTdistAlanine}
\end{figure}

Molecular dynamics (MD) simulations are very powerful, providing microscopic insights into the mechanisms underlying physical and chemical condensed phase processes. However, due to their atomic spatial and temporal resolution, standard MD simulations are limited to events that occur on timescales shorter than $\sim 1 \, \mu s$ \cite{barducci_metadynamics_2011,yang_thermodynamics_2015}.
In many cases, the complex dynamics of the system lead to longer timescales, through a very broad distribution of transition times between metastable states, also known as first-passage times \cite{salvalaglio_assessing_2014} (FPT).
To demonstrate this, Fig.~\ref{fig:FPTdistAlanine} presents the probability density, denoted by $f \left( \tau \right)$, of the FPT, $\tau_1, \tau_2, ..., \tau_N$, obtained from $N$ simulations of transitions between the two conformers of an alanine dipeptide molecule -- a common model system \cite{salvalaglio_assessing_2014,tiwary_metadynamics_2013}.
It shows that many transitions occur on a timescale much shorter than $1 \, \mu s$ -- more than 25\% of them under $100 \, ns$. However, the tail of the distribution decays so slowly that the mean FPT is almost an order of magnitude larger,  $759 \, ns$, and some trajectories fail to complete even after $4 \, \mu s$. 
There is thus an ongoing effort to develop procedures for expediting such processes \cite{yang_enhanced_2019, henin_enhanced_2022}.

Stochastic resetting (SR) is the procedure of occasionally stopping and restarting random processes using independent and identically distributed initial conditions. 
The resetting times are typically taken at constant intervals (“sharp resetting”) or from an exponential distribution with a fixed rate (“Poisson resetting”). The interest in SR has grown significantly since the pioneering work of Evans and Majumdar \cite{evans_diffusion_2011}. 
They showed that while a particle undergoing Brownian motion between two fixed points in space has an infinite mean FPT, its mean FPT with SR becomes finite. Therefore, the particle reaches the target point infinitely faster on average. This result has effectively established an emerging field of research in statistical physics, to which a recent special issue was dedicated.\cite{evans_stochastic_2020,kundu_stochastic_2022}

The power of resetting in accelerating random processes has been widely demonstrated: in randomized computer algorithms \cite{luby_optimal_1993,gomes_boosting_1998,montanari_optimizing_2002}, in various search processes \cite{kusmierz_optimal_2015,bhat_stochastic_2016,chechkin_random_2018,ray_peclet_2019,robin_random_2019,evans_run_2018,pal_search_2020,bodrova_resetting_2020}, experimentally in systems of colloidal particles \cite{tal-friedman_experimental_2020,besga_optimal_2020}, in queuing systems \cite{bressloff_queueing_2020,bonomo_mitigating_2022}, and in the Michaelis–Menten model of enzymatic catalysis, where resetting occurs naturally by virtue of enzyme-substrate unbinding \cite{reuveni_role_2014,rotbart_michaelis-menten_2015}. The latter finding was then leveraged to develop a general treatment of first-passage processes under restart \cite{pal_first_2017}. There, it was shown that the FPT distribution in the absence of SR can be used to determine the FPT distribution with resetting. Moreover, the mean and standard deviation of the FPT distribution without resetting are enough to determine a sufficient condition for SR to expedite a random process\cite{pal_inspection_2022}. Specifically, if the ratio of the standard deviation to the mean FPT (the coefficient of variation, COV) is greater than one, a small reset rate $r$ is guaranteed to lower the mean FPT. The slowly-decaying distributions that occur in molecular simulations of long-timescale processes can also have a COV that is greater than one. For example, the distribution in Fig.~\ref{fig:FPTdistAlanine} has a COV of $\sim 1.3$. This indicates that resetting can expedite MD simulations.

In this work, we use SR for the first time for enhanced sampling of molecular simulations.
MD simulations are an exciting playground for the application of resetting, while raising new fundamental questions that are of interest to both communities. 
In SR, the unbiased kinetics (without resetting) are known, and the goal is to understand how much speedup can be gained by restarting the random process.
On the other hand, in the MD community, the long-timescale processes cannot be accessed directly and enhanced sampling methods are required to expedite them. Introducing SR for this purpose raises the question of inference -- can we obtain the free energy surfaces and the kinetics of reset-free processes from simulations with SR?
This question has not been explored in the SR community but is the natural goal of enhanced sampling methods.

Various methods have been developed in the field of molecular simulations to overcome the long-timescale problem, such as umbrella sampling \cite{torrie_nonphysical_1977,kastner_umbrella_2011}, 
Metadynamics \cite{valsson_enhancing_2016,barducci_metadynamics_2011,sutto_new_2012,bussi_using_2020}, 
on-the-fly probability enhanced sampling (OPES) \cite{invernizzi_rethinking_2020,invernizzi_unified_2020,invernizzi_opes_2021}, 
and adiabatic free energy dynamics \cite{abrams_efficient_2008,rosso_adiabatic_2002,rosso_use_2002}. 
Many of them rely on identifying suitable collective variables -- effective reaction coordinates that ideally describe the slowest modes of the process \cite{invernizzi_making_2019}. 
Below, we show that SR can be used for enhanced sampling without finding suitable collective variables, which is highly non-trivial for condensed phase processes \cite{sidky_machine_2020,chen_collective_2021}.
Most importantly, we demonstrate that the mean transition times without resetting, that are often too long to be sampled directly, can be recovered from accelerated simulations performed at a single restart rate. In this letter we give a proof of concept for these desirable features using examples ranging from simple models to a molecular system. We obtain a speedup by an order of magnitude in some cases. Our method opens new avenues in both the MD and SR communities, hopefully promoting a fruitful collaboration between the two.

\section{Results and discussion}

\begin{figure}
  \includegraphics[width=\linewidth]{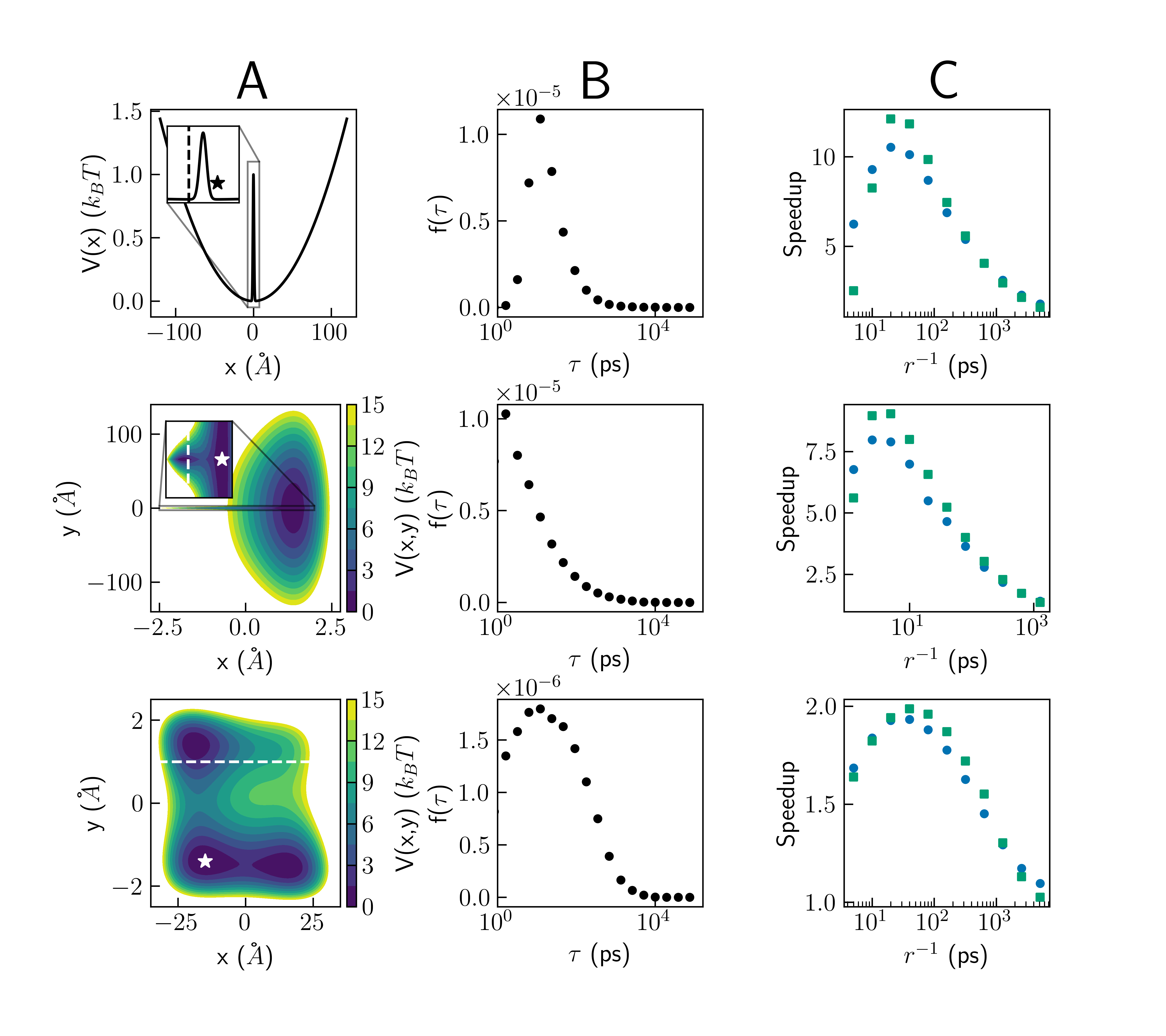}
  \caption{The potential energy surface (column A), the FPT distribution without resetting (column B) and the speedups obtained using Poisson (blue circles) and sharp (green squares) resetting (column C) for the one dimensional double-well model (top row), the model of Gimondi et al.\cite{gimondi_building_2018}  (middle row) and the modified Wolfe-Quapp potential (bottom row). The full potential details are given in the SI.}
  \label{fig:modelPotentialsAndSpeedup}
\end{figure}

We begin by demonstrating that SR can indeed enhance the sampling of MD simulations.
Mathematically, we know that if the COV is greater than one, it is guaranteed that resetting can expedite the process.
But for what potential energy surfaces do we expect this to occur?
We answer this question using three illustrative model systems representing possible scenarios in MD simulations.
Resetting was successful in accelerating transitions in all of them, and, for two of them, we obtained an order of magnitude speedup in the mean FPT. To benchmark our approach, we chose the parameters of the model potentials such that the mean FPT without resetting is accessible ($\sim 1 \, ns)$ to allow extensive sampling of the unbiased process.
Below, we briefly describe the models while the full parameters are given in the SI. The results for each model are given in a separate row in Fig.~\ref{fig:modelPotentialsAndSpeedup}. In all cases, the left panel shows the potential and the middle panel presents the FPT probability density $f(\tau)$ without resetting. The right panel shows the speedup obtained by both Poisson and sharp resetting, at different restart rates $r$. All simulations are of a single particle initialized at fixed positions, denoted by stars in the left panels of Fig.~\ref{fig:modelPotentialsAndSpeedup}, with an initial velocity sampled from the Maxwell-Boltzmann distribution at $300 \, K$. The dashed line in Fig.~\ref{fig:modelPotentialsAndSpeedup} defines the spatial threshold for the first passage. The simulations were performed in the Large-scale Atomic/Molecular Massively Parallel Simulator (LAMMPS) \cite{LAMMPS}, with SR easily implemented in the input files. Full details and input examples are given in the SI and the corresponding GitHub repository \cite{Blumer_Input_files_for}.

The first model is presented in the top row of Fig.~\ref{fig:modelPotentialsAndSpeedup}. It is a one dimensional double-well potential that is composed of a trapping harmonic term and a Gaussian centered at $x=0 \, \AA$. The model has two symmetric minima that are separated by a moderate barrier ($1 \, k_BT$). The harmonic spring constant was taken to be soft, such that the particle can explore areas very far away from the center ($\sim 100 \, \AA$). This model, with a different choice of parameters, was previously used to describe the umbrella inversion in ammonia~\cite{swalen_potential_2004}. The simulations were initiated at the right minimum ($x=3 \, \AA$) and the FPT was defined as reaching the second basin ($x \le -3 \, \AA$).
The distribution without resetting is broad, spanning about four orders of magnitude (note the logarithmic timescale), and has a COV of {$\sim 2.9$}. In the absence of resetting, some transitions occur as fast as a few picoseconds while others take as long as tens of nanoseconds. The median FPT is {$125 \, ps$} and the mean FPT is {$1325 \, ps$}. By introducing SR, we were able to reduce the mean FPT by more than an order of magnitude, with a speedup of $10.5$ and $12.1$ for Poisson and sharp resetting, respectively. The results agree with previous work showing that sharp resetting is guaranteed to lead to higher optimal speedups than any other resetting protocol \cite{pal_first_2017}.

The second model is presented in the middle row of Fig.~\ref{fig:modelPotentialsAndSpeedup}. It is a two dimensional potential, introduced by Gimondi et al.\cite{gimondi_building_2018} (with slightly different parameters) to represent two isoenergetic states with very different contributions to the entropy. It has two basins located at $ \left( x=\pm 1.3 \, , y=0 \right)\, \AA$, which are separated by a barrier of $\sim 3 \, k_BT$ centered at the origin. Note that the left basin is so narrow it can only be clearly seen in the figure inset. The basins have the same width in the x-direction, but in the y-direction, the right one is much broader ($\sim 50 \, \AA$) than the left one ($\sim 0.5 \, \AA$).
As a consequence, the particle can freely explore areas in the right basin where it cannot cross to the other well. The simulations were initiated from the right basin, and the FPT was defined as crossing to the left well ($x \le -1 \, \AA$).
The results are similar to those of the one dimensional model. The unbiased FPT distribution is broad, with values ranging from $1 \, ps$ to $20 \, ns$. The median and mean of the distribution are $450\, ps$ and $1125 \, ps$, respectively. The COV is smaller than the one found for the double-well example ($1.44$), but the speedup is similar -- $8.0$ for Poisson resetting and $9.0$ for sharp resetting.

The final model system is presented in the bottom row of Fig.~\ref{fig:modelPotentialsAndSpeedup}. It is a modified version of the Wolfe-Quapp potential, often used for benchmarking enhanced sampling methods.\cite{invernizzi_making_2019,quapp_growing_2005,ray_rare_2022}. This potential has two metastable basins, one at $y<0$ and the other at $y>0$. The former is divided into two sub-states 
that have similar width and depth. 
The lower sub-states are $30 \, \AA$ apart and are separated by a moderate barrier ($\sim 1.5 \, k_BT$).
Larger barriers separate the lower basin from the upper well, $\sim 6.25 \, k_BT$ and $\sim 10 \, k_BT$ for the left and right lower sub-states, respectively. This makes the transition to the upper well much more probable from the lower left sub-state than the right one. Therefore, this model is an example of a system in which the particle can either cross to the upper well, completing the process, or spend long periods of time in a less reactive nearly isoenergetic state.
The simulations were initialized in the lower left sub-state $\left( x=-14.9, y=-1.4\right)\,\AA$ and the FPT was defined as crossing to the upper basin, $y \ge 1 \, \AA$.
The obtained FPT distribution without resetting is again very broad, spanning from a few picoseconds for the fastest transitions to tens of nanoseconds for the slowest. 
We find that, while this model has a very similar COV, mean and median FPT as the second example above (1.43, $1125 \, ps$ and $500 \, ps$ respectively), the obtained speedup is smaller, $\sim2$ for both sharp and Poisson resetting.
This is because the modified Wolfe-Quapp potential has a mean FPT that is only two orders of magnitude larger than the most probable value, compared to three orders of magnitude in the previous example.
This result shows that while a COV greater than one guarantees that SR would accelerate the process, the entire shape of the unbiased FPT distribution determines the resulting speedup. In this context, we note a recent development by Starkov and Belan~\cite{starkov_universal_2022}. 

\begin{figure}
  \includegraphics[width=0.5\linewidth]{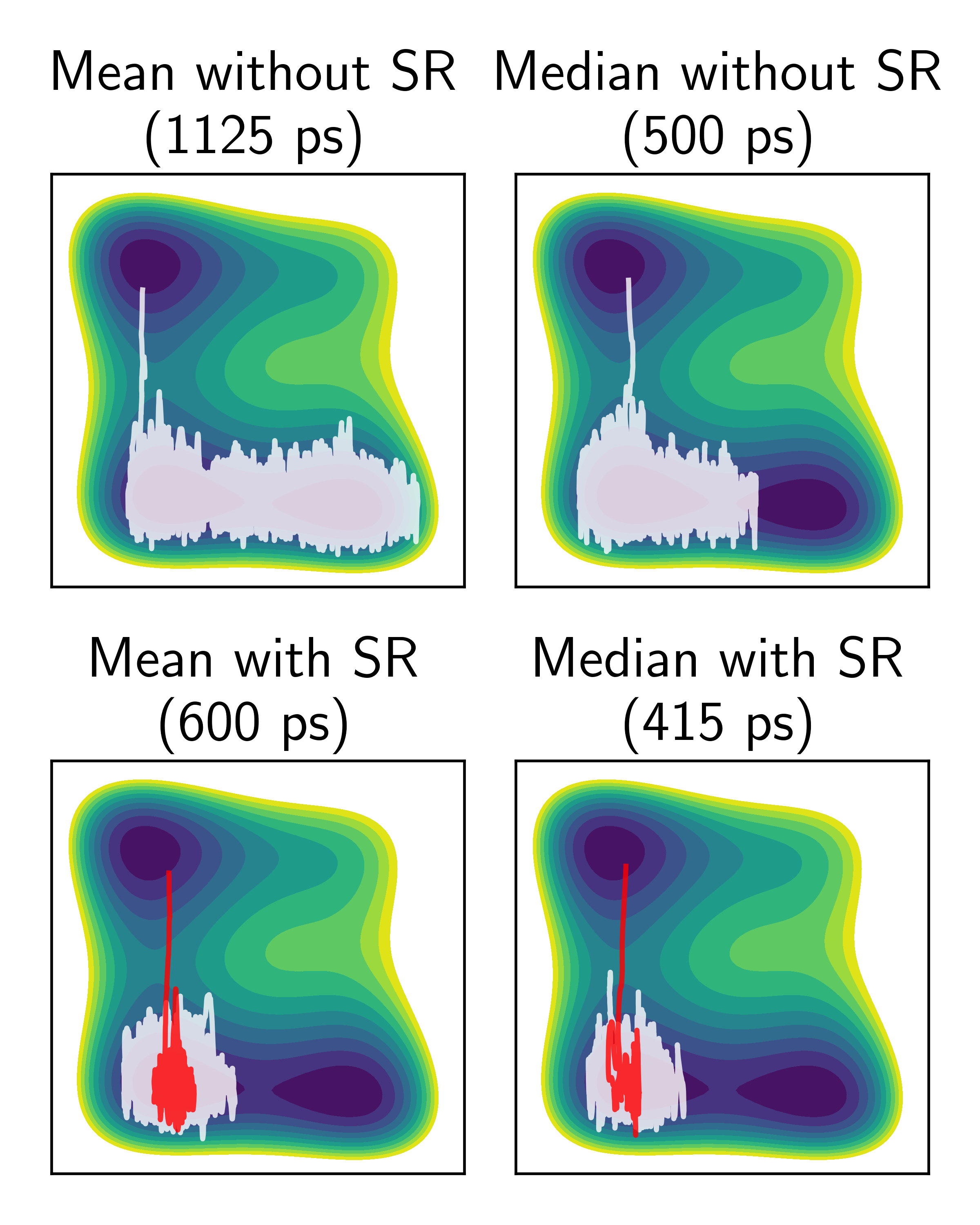}
  \caption{Selected trajectories with FPT of different timescales, without resetting (top row) and with sharp resetting every $40 \, ps$ (bottom row). The full trajectories are presented in white. For the trajectories with SR, the last leg following the final reset event and until the crossing of the barrier is highlighted in red.}
  \label{fig:trajectories}
\end{figure}

It is interesting to test whether SR affects the transition paths between metastable states. We have checked this in Fig.~\ref{fig:trajectories}, plotting trajectories for the modified Wolfe-Quapp potential with transition times representing the mean and median of the FPT distributions with and without resetting. It can be seen that both trajectories with resetting stay localized in the lower-left basin before crossing to the upper well while the trajectories without resetting explore a much broader area of the lower basin, spending more time in nonreactive configurations. The lower panels also show in red the part of the simulations between the last restart and the crossing to the upper well. We find that the final leg of the trajectory shows a similar distribution of transition paths as in the simulations without resetting. This is because SR does not change the dynamics between restart events, unlike other biasing algorithms that continuously add energy to the system\cite{bussi_using_2020, valsson_enhancing_2016}, which may result in transitions through highly unlikely paths.

Finally, to demonstrate that SR can be a useful tool in more realistic molecular simulations, we also applied it to accelerate a classic example of enhanced sampling -- the alanine dipeptide molecule. It has two states, usually referred to as the $C_{7eq}$ and $C_{7ax}$ conformers~\cite{valsson_enhancing_2016}, which differ by their values of two dihedral angles, $\phi$ and $\psi$ (see Fig.~\ref{fig:FPTdistAlanine}(a)). The simulations were initiated from the more stable $C_{7eq}$ conformer after energy minimization, for which $\phi < 0 \, \text{rad}$, and the FPT was defined by $ 0 \le \phi \le 2 \, \text{rad}$. To the best of our knowledge, this is the first application of SR to a molecular system. 

Going beyond the mean FPT, we compare the full distributions with and without resetting in Fig.~\ref{fig:FPTdistAlanine}(b). Our results shed light on how SR leads to acceleration. It effectively eliminates transition times that are much longer than $1/r$, leading to a narrower distribution. A speedup of $2.3$ is obtained, reducing the mean FPT from $759 \, ns$ without resetting to $333 \, ns$ with SR. We find that the speedup is not very sensitive to the resetting rates used from $0.1 \, ns^{-1}$ to $0.01 \, ns^{-1}$ for this system. 
In such a well-studied model, with known efficient collective variables, methods such as Metadynamics or OPES admittedly result in much higher speedups. However, identifying suitable collective variables in condensed phases is still generally very challenging. The great appeal of SR is that no collective variables are needed and only very minimal prior knowledge on the timescales without resetting is required. Moreover, SR can be used in a complimentary fashion to Metadynamics or OPES. These simulations are usually performed with suboptimal collective variables in practice~\cite{invernizzi_making_2019}. If their COV is greater than one, introducing SR will lead to further speedup.

To conclude the first part of this Letter, we showed that SR is able to expedite transitions in MD simulations ranging from simple models to a molecular system, with up to an order of magnitude reduction of the mean FPT. We examined the sensitivity of the results to the definition of the FPT and the initial conditions (e.g., to sampling the initial position from a distribution). Our findings did not change significantly and, in some cases, the speedups obtained were even greater. See the SI for a detailed discussion.

Accelerating transitions between metastable states is very useful, as it can be used to generate data for training neural network potential energy surfaces~\cite{bonati_silicon_2018}, to identify collective variables~\cite{sidky_machine_2020}, and to predict previously undiscovered intermediates~\cite{piaggi_predicting_2018}.
Next, we tackle another major goal of enhanced sampling -- the inference of the unbiased kinetics from biased simulations. Despite many recent advancements \cite{palacio-rodriguez_transition_2022,salvalaglio_assessing_2014,tiwary_metadynamics_2013,ray_rare_2022,mandelli_metadynamics_2020}, evaluating the rates of long-timescale processes from enhanced simulations is still very challenging, and they can deviate by orders of magnitude from experiments \cite{blow_seven_2021}. To increase the accuracy, methods such as infrequent Metadynamics or OPES-flooding use much weaker biasing~\cite{tiwary_metadynamics_2013, salvalaglio_assessing_2014, ray_rare_2022, palacio-rodriguez_transition_2022}, and the resulting speedups are significantly lower than standard Metadynamics.
Here, we employ SR for this purpose, showing that it is not limited to expediting transitions, but can also be used for inferring kinetics.
We now explain how to obtain the mean FPT without resetting using data from accelerated trajectories at a \textit{single} restart rate.  

\begin{figure}
  \includegraphics[width=\linewidth]{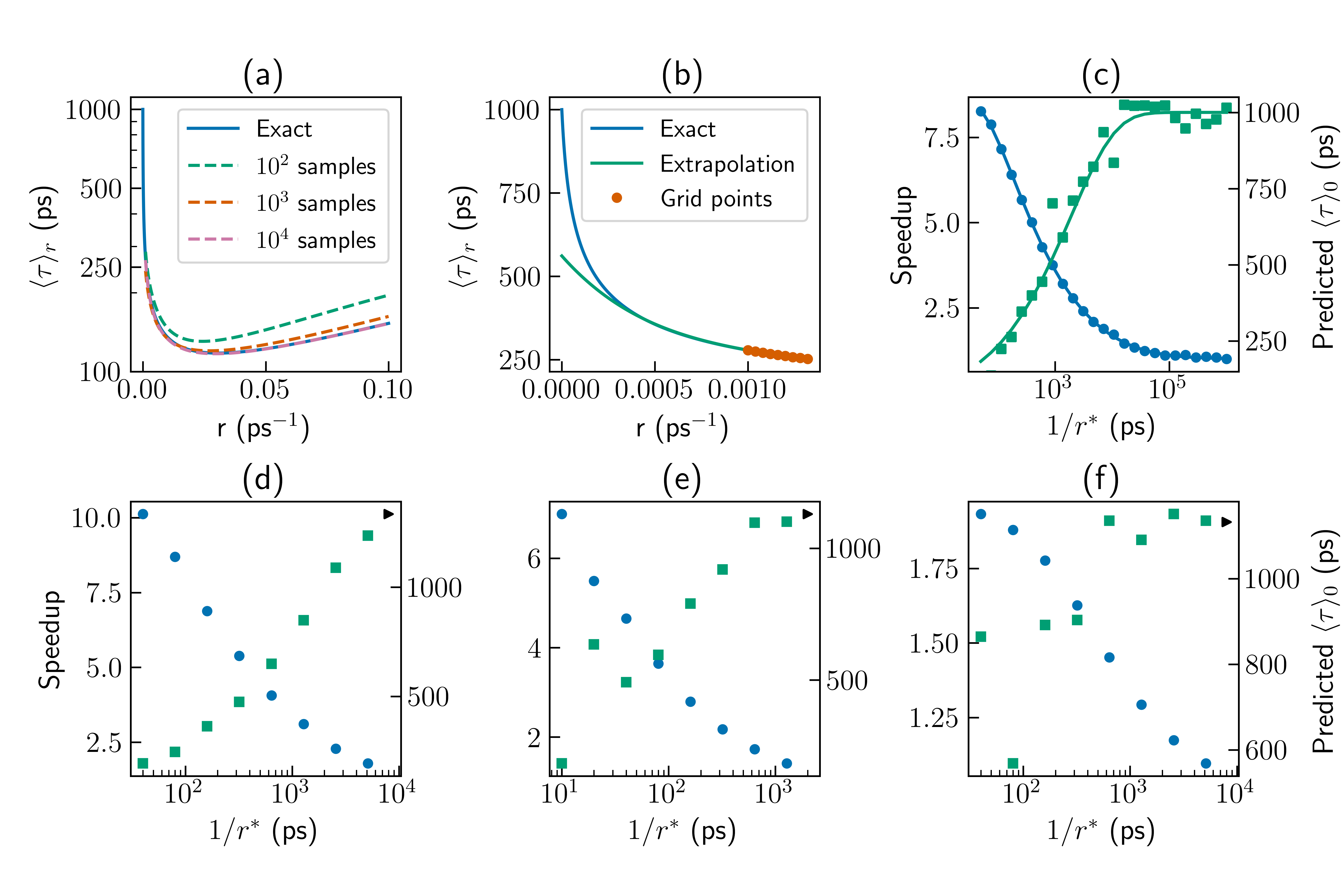}
  \caption{Top row: Results for an inverse Gaussian distribution with an unbiased mean FPT of $1000 \, ps$ (see SI for details). (a) Exact $\langle \tau \rangle_r$ obtained by using the analytic Laplace transform in Eq.~\ref{eqn:LaplaceTransformAndFPT} and approximate values using a different number of trajectories at reset rate $r^*=0.001 \, ps^{-1}$ to evaluate it numerically. (b) Exact $\langle \tau \rangle_r$ and its fourth order Taylor series around $r^*=0.001 \, ps^{-1}$ using the indicated grid points (c) Speedup (blue circles) and $\langle \tau \rangle_{0}$ predictions (green squares), obtained by extrapolation of the Taylor series to $r=0$, as a function of $1/r^*$. 
  In panel (c), lines represent predictions using the analytical Laplace transform, while the dots show the results using 50,000 trajectories in the evaluation of the numerical Laplace transform.
  Bottom row: Speedup (blue circles) and $\langle \tau \rangle_{0}$ predictions (green squares) against $1/r^*$ for (d) the one-dimensional double-well potential, (e) the potential introduced by Gimondi et. al., (f) The modified Wolfe-Quapp potential. The black arrows indicate the $\langle \tau \rangle_{0}$ obtained in unbiased simulations.
  }
  \label{fig:analyticPredictionAndExtrapolation}
\end{figure}

For long-timescale processes ($> 1 \, \mu s$)  we cannot determine the FPT distribution without resetting. Instead, we can accelerate the simulations and obtain the mean FPT at several reset rates $r>0$. It is then possible to extrapolate the results to the $r=0$ limit to get an estimate of the unbiased mean FPT. However, this is a very expensive procedure, since typically thousands of transitions are required to converge the FPT distributions and the reset rate that leads to optimal speedup is unknown a priori. Fortunately, we find that for Poisson resetting the FPT distribution at any reset rate $r^*$, denoted by  $f_{r^*}(\tau)$, is enough to predict the mean FPT, $\langle \tau \rangle_r$, at all $r>r^*$ through
\begin{equation}
  \langle \tau \rangle_r = \frac{1-\tilde{f}_{r^*}(r-r^*)}{\left(r-r^*\right)\tilde{f}_{r^*}(r-r^*)}, \label{eqn:LaplaceTransformAndFPT}
\end{equation}
where the Laplace transform of $f_{r^*}(\tau)$ is defined as
\begin{equation}
  \tilde{f}_{r^*}(s) = \int_0^\infty e^{- s \tau} f_{r^*} \left( \tau \right) \, \mathrm{d}\tau = \langle e^{-s \tau } \rangle_{r^*}.
\end{equation}
Eq.~\ref{eqn:LaplaceTransformAndFPT} is exact, given we have the Laplace transform, and its derivation is given in the SI.
In practice, we evaluate the Laplace transform by performing $N$ simulations at a single reset rate $r^*$. We determine $\tilde{f}_{r^*}(r-r^*) $ for a set of discrete values $r>r^*$ by taking the arithmetic mean of $e^{-(r-r^*)\tau_j}$, where $\tau_j$ is the FPT of the j-th trajectory. Then, we use Eq.~\ref{eqn:LaplaceTransformAndFPT} to predict the mean FPT for the selected values of $r>r^*$. 
We verify this procedure in Fig.~\ref{fig:analyticPredictionAndExtrapolation} for an inverse Gaussian FPT distribution, whose Laplace transform is known analytically. This distribution describes the FPT of drift diffusion to an absorbing boundary~\cite{folks_inverse_1978}. The full details of the simulations to determine the Laplace transform at reset rate $r^*$ numerically are given in the SI. 
Panel (a) shows that using ten thousand samples to evaluate the Laplace transform numerically is sufficient to reproduce the result obtained using the analytical transform very accurately.

Finally, using the values of $\langle \tau \rangle_r $ predicted from simulations at a single reset rate $r^*$ we can extrapolate to $r=0$ and get the unbiased mean FPT at a much lower cost than directly performing simulations at many reset rates.
Fig.~\ref{fig:analyticPredictionAndExtrapolation} (b) demonstrates the extrapolation procedure. It is based on predicting $\langle \tau \rangle_r$ on a grid of points in the vicinity of $r^*$ and fitting them with a fourth order Taylor series. The mean FPT without resetting is then obtained from the value of the fitted function at $r=0$. We compared several extrapolation approaches, which resulted in similar accuracy.  See the SI for a full comparison.
Fig. \ref{fig:analyticPredictionAndExtrapolation} (c) shows the predicted unbiased mean FPT, $\langle \tau \rangle_0$, as a function of $1/{r^*}$.
Naturally, the estimation of the unbiased mean FPT from the extrapolation becomes exact as $r^*$ goes to zero. However, the speedup also decreases in this limit. This results in a trade-off between precision and speedup. A similar trade-off was also observed by Ray et. al. for the OPES flooding enhanced sampling method \cite{ray_rare_2022}. For this benchmark, we obtained an error of $\sim10\%$ in the prediction of the unbiased mean FPT for a speedup of $\sim1.7$, an error of $\sim50\%$ for a speedup of $\sim2.8$, an error of $\sim100\%$ for a speedup of $\sim 3.9$ and an error of $\sim500\%$ for a speedup of $\sim 8.0$. Also in the case of inference, the strength of SR is that it does not require identifying efficient collective variables. While the speedup and accuracy of the kinetic information obtained from other enhanced sampling methods is sensitive to the collective variables used\cite{ray_rare_2022}, resetting has a single parameter -- the restart rate -- that can be tuned to control the balance between accuracy and speedup.

We have also predicted the unbiased FPT by the same method for the model potentials above. Results are given in panels (d)-(f) of Fig. \ref{fig:analyticPredictionAndExtrapolation} as was presented for the inverse Gaussian distribution in Panel (c). For the one-dimensional model (d) we obtained an error of $\sim3\%$ for a speedup of $\sim1.7$, an error of $\sim45\%$ for a speedup of $\sim2.8$, an error of $\sim100\%$ for a speedup of $\sim 4.1$ and an error of $\sim595\%$ for a speedup of $\sim10.1$. Similarly, in the second model system (e), we obtained an error of $\sim8\%$ for a speedup of $\sim1.8$, an error of $\sim55\%$ for a speedup of $\sim3.1$, an error of $\sim90\%$ for a speedup of $\sim 3.6$ and an error of $\sim515\%$ for a speedup of $\sim7.0$. For the modified Wolfe-Quapp potential (f), we obtained an error of $\sim2\%$ for a speedup of $\sim1.4$ and an error of $\sim30\%$ for a speedup of $\sim 1.9$.

Finally, Eq.~\ref{eqn:LaplaceTransformAndFPT} can also be used to find the reset rate which gives the maximal speedup at almost no cost. This is shown in Panel (a) of Fig.~\ref{fig:analyticPredictionAndExtrapolation}, in which we tested the sensitivity of the prediction of Eq.~\ref{eqn:LaplaceTransformAndFPT} to the number of trajectories used. It can be seen that as little as a hundred samples lead to predictions that capture the qualitative behavior of the mean FPT as a function of the reset rate. While this is not sufficient statistics for the inference of unbiased kinetics, it gives a good estimate for the optimal reset rate and speedup.

\section{Conclusions}



To conclude, we employed stochastic resetting to enhance the sampling of long-timescale processes in molecular dynamics simulations for the first time. In applications ranging from toy models to a molecular system, we obtained speedups of up to an order of magnitude in the mean first-passage time. 
The most appealing feature of stochastic resetting as an enhanced sampling method is its incredible simplicity -- just restart the simulations at random times to accelerate them. No collective variables are required, and only a coarse estimate of a reset rate that would result in speedup is needed.
The optimal speedup can then be predicted through Eq.~\ref{eqn:LaplaceTransformAndFPT}.
We demonstrated the usefulness of stochastic resetting as a standalone approach to enhance the sampling of MD simulations, but resetting can also be combined with existing algorithms, such as Metadynamics, to further accelerate simulations performed with suboptimal collective variables (given a COV $> 1$). It will be exciting to attempt such a combination on larger and more complex condensed phase systems in the near future. 


We also showed that simulations at a \textit{single} reset rate $r^*$ are enough to infer the mean first-passage time without resetting with adequate accuracy. This is achieved by combining forward prediction to $r > r^*$, via Eq.~\ref{eqn:LaplaceTransformAndFPT}, with backward extrapolation to $r=0$. 
In doing so, we have brought inference in stochastic resetting to the foreground, setting the stage for future theoretical developments.
Our method opens new avenues in both the molecular dynamics and stochastic resetting communities, hopefully promoting a fruitful collaboration between the two.
\begin{acknowledgement}

Barak Hirshberg acknowledges support by the USA-Israel Binational Science Foundation (grant No. 2020083) and the Israel Science Foundation (grants No. 1037/22 and 1312/22). Shlomi Reuveni acknowledges support from the Israel Science Foundation (grant No. 394/19). This project has received funding from the European Research Council (ERC) under the European Union’s Horizon 2020 research and innovation program (grant agreement No. 947731).

\end{acknowledgement}


\bibliography{main}


\end{document}




\section{General simulation details}
All molecular dynamics (MD) simulations were performed in the Large-scale Atomic/Molecular Massively Parallel Simulator (LAMMPS) \cite{LAMMPS}. 
The simulations of the one- and two-dimensional models were performed in the canonical (NVT) ensemble at a temperature $T=300 \, K$, using a Langevin thermostat with a friction coefficient $\gamma=0.01 \, fs^{-1}$. The integration time step was $1 \, fs$. We simulated a single particle with mass $m=40 \, g \, mol^{-1}$, representing an argon atom.
$50000$ trajectories were sampled for every model system presented in the main text or in the SI. We checked whether a transition occurred every $1 \,ps$ for simulations without resetting, and every $0.1 \,ps$ for all simulations with resetting except those with rate $r=8 \,ps^{-1}$, for which we checked every $0.05 \, ps$.

The simulations of an isolated alanine dipeptide molecule were performed at $300 \, K$ using a time step of $2 \, fs$ with a Nos\'e-Hoover chains thermostat \cite{martyna_nosehoover_1992} and a temperature damping parameter of $100 \, \Delta t$. The condition for the FPT was checked every 100 time steps. We used the AMBER99SB force field. Available GROMACS input files for this system \cite{noauthor_plumed_nodate} were converted to LAMMPS format using Intermol \cite{shirts_lessons_2017}. With them, we obtain a free energy difference between the conformers of $\sim 9 \, kJ \, mol^{-1}$, which is in reasonable agreement with reference values \cite{bonati_data-driven_2020}. The FPT distributions with and without resetting were obtained from 10000 trajectories each. For the trajectories without resetting, 307 trajectories did not show a transition within $4 \, \mu s$. They are not plotted in the probability density of Fig. 1 but are included in calculating the mean FPT and speedup. We note that, as  a result, the speedup gained by resetting that we report for alanine dipeptide is a lower bound.

\section{Implementation of stochastic resetting}
Stochastic resetting (SR) was implemented in the input files as explained below. A stopping mechanism after the first transition was also incorporated through the LAMMPS input. The initial velocities, and their values after each reset event, were sampled from the Maxwell-Boltzmann distribution at the relevant temperatures using Python. For Poisson resetting, waiting times between resets were also sampled using Python, from an exponential distribution, $f\left(\tau\right)=re^{-r\tau}$, where $r$ is the restart rate.
Below, we give a simplified example of the implementation for a two-dimensional simulation with only three reset events for clarity. The initial position in this example is fixed at $\left(1, 0\right) \AA$ and the first transition (passage) is defined as crossing $x=0 \, \AA$. Full example input files are given in the corresponding GitHub repository \cite{Blumer_Input_files_for}.

\begin{verbatim}
variable resetTimes index 9 133 22 # Waiting times between resets in ps
variable initialX equal 1
variable initialY equal 0
variable initialVx index -0.00113 0.00278 0.00650
variable initialVy index 0.00120 0.00233 -0.000394
variable reactionCoordinate equal "x[1]" # The x coordinate of the particle
variable passageCriterion equal 0

label mainLoop
variable a loop 3 # Loop over total number of reset events

label innerLoop
variable b loop ${resetTimes}
run 1000 # Run for 1ps and check whether a transition occurred
if "(${reactionCoordinate} < ${passageCriterion})" then &
   "jump SELF break"
next b
jump SELF innerLoop

set atom 1 x ${initialX} y ${initialX} vx ${initialVx} vy ${initialVy} # Reset...
next a
next Vx
next Vy
next resetTimes
jump SELF mainLoop

label break
print "ALL DONE"
\end{verbatim}

\section{Laplace transforms for the inverse Gaussian distribution}

We used Eq.~\ref{eqn:InverseGaussian} for the inverse Gaussian probability density function. The expression for its analytical Laplace transform is given in Eq.~\ref{eqn:InverseGaussianLaplace}~\cite{}.
We used the parameters $L = 1000$, $V = 1 \, ps^{-1}$ and $D = 12500\, ps^{-1}$, which lead to a mean FPT of $1 \, ns$ and a coefficient of variation (COV) of 5.
In the context of drift diffusion, $L$ is the initial distance from the boundary, $V$ is the drift velocity and $D$ is the diffusion constant.

\begin{equation}
f(\tau) = \frac{L}{\sqrt{4\pi D \tau^3}} \exp \left(-\frac{(L - V\tau)^2}{4D\tau}\right)  \label{eqn:InverseGaussian}
\end{equation}

\begin{equation}
\tilde{f}(s) = \exp \left(\frac{L}{2D}\cdot \left(V-\sqrt{V^2+4Ds}\right)\right)  \label{eqn:InverseGaussianLaplace}
\end{equation}

To evaluate the Laplace transform of the inverse Gaussian distribution numerically, simulations of first-passage times (FPT) were performed in Python using Scipy \cite{2020SciPy-NMeth} . Trajectories at a reset rate $r^*$ were obtained in the following way: At each step, we sampled a new passage time $\tau_{passage}$ and a new reset time $\tau_{reset}$ from their corresponding distributions.
If we found $\tau_{passage}<\tau_{reset}$, it meant that there has been a passage before the next reset time. $\tau_{passage}$ was added to the overall simulation time and the simulation was stopped. Otherwise, it meant that the process was restarted before a transition occurred. $\tau_{reset}$ was then added to the overall simulation time and we proceeded to sample new values of $\tau_{reset}$ and $\tau_{passage}$. We continued this procedure until we encountered a successful passage, $\tau_{passage}<\tau_{reset}$. 

\section{Model potentials}
\label{sec:ModelPotentials}
Here we present the exact equations and parameters of the chosen model potentials. The parameters are given such that spatial distances are in $\AA$ and potential energies are in units of $1 \, k_BT$ for a temperature of $300 \, K$.

The one dimensional double-well is described by Eq. \ref{eqn:1Dpotential}, with $A=\num{1e-4}$, $B=1$, $C=1$.

\begin{equation}
 V(x) = Ax^2 + B \exp \left(-Cx^2\right)
 \label{eqn:1Dpotential}
\end{equation}

The form of the two dimensional potential introduced by Gimondi et al.\cite{gimondi_building_2018} is given in Eq. \ref{eqn:2Dpotential}. Most of the parameters are taken as chosen there:
$x_1=2.5$, $x_2=-2.5$, $\sigma_1=1.3$, $\sigma_2=1.3$, $y_1=0$, $y_2=0$, $\lambda_2=1$. To make the right basin  broader, we used a larger value of $\lambda_1=2000$, and a smaller coefficient for the y-coordinate harmonic spring. In order to achieve an accessible mean FPT in the absence of resetting, we lowered the barrier and chose $A_1=A_2=41$.

\begin{equation}
V(x,y)=-\sum_{i=1}^2A_i \exp \left(-\frac{\left(x-x_i\right)^2}{2\sigma_i^2}-\frac{\left(y-y_i\right)^2}{2\lambda_i^2}\right)+4x^2+\num{5e-4}y^2 \label{eqn:2Dpotential}
\end{equation}

The modified Wolfe-Quapp potential is of the form given in Eq. \ref{eqn:WolfeQuapp}. We modified the original potential \cite{quapp_growing_2005} by replacing the x coordinate with a rescaled $x' = x / 15$, increasing the coefficients of the linear terms in both coordinates, and multiplying the resulting potential by a factor of 1.5. These modifications were done in order to achieve two remote, distinct sub-states with similar stability in the lower basin.

\begin{equation}
V(x,y) = 1.5\left(x'^4+y^4-2x'^2-4y^4+1.5x'+1.2y+x'y\right)
\label{eqn:WolfeQuapp}
\end{equation}

\section{Derivation of Eq. 1 of the main text}

Here, we will derive Eq. 1 from the main text, which connects the mean FPT at reset rates $r$, $\langle \tau \rangle_r$, to the FPT distribution at some reset rate $r^*$, denoted by  $f_{r^*}(\tau)$.
We begin with Eq.~\ref{eqn:originalLaplaceTransformAndFPT}, derived by 
Reuveni,~\cite{PhysRevLett.116.170601} which connects the FPT distribution without resetting of a random process, $f(\tau)$, to its mean FPT with Poisson resetting at rate $r$, through 
\begin{equation}
  \langle \tau \rangle_r = \frac{1-\tilde{f}(r)}{r\tilde{f}(r)}, \label{eqn:originalLaplaceTransformAndFPT}
\end{equation}
which is identical to Eq. 1 of the main text for $r^*=0$. This equation holds for any distribution $f(\tau)$, including the special case $f(\tau)=f_{r^*}(\tau)$. Consequently, we may treat the process with reset rate $r^*$ as if it is an \textit{unbiased} random process and ask what happens when adding a resetting procedure with rate $r'$ to it. We  rewrite the equation above for this special case in Eq. \ref{eqn:modifiedLaplaceTransformAndFPT}. The new notation emphasizes that $r'$ is independent of $r^*$ and receives any values $r'\geq 0$, as opposed to $r$ in Eq. 1 of the main text, which only receives $r\geq r^*$. We signify the two independent resetting procedures by two subscripts in the left-hand side of the equation.

\begin{equation}
  \langle \tau \rangle_{r^*,r'} = \frac{1-\tilde{f}_{r^*}(r')}{r'\tilde{f}_{r^*}(r')} \label{eqn:modifiedLaplaceTransformAndFPT}
\end{equation}

What is the distribution of the resulting resetting times? We combined two resetting procedures, each an individual Poisson process with reset times sampled from an exponential distribution, with a rate $r^*$ or $r'$.
Due to the additive property of Poisson processes, this results in another Poisson process, with rate $r=r^*+r'$~\cite{kingman_poisson_1993}. Thus, the combined effect is equivalent to the introduction of a single resetting rate $r = r^* + r'$, meaning $\langle \tau \rangle_{r^*,r'}=\langle \tau \rangle_{r^*+r'}=\langle \tau \rangle_r$. Substituting this fact into Eq. \ref{eqn:modifiedLaplaceTransformAndFPT} yields Eq. 1 of the main text,

\begin{equation}
  \langle \tau \rangle_r = \frac{1-\tilde{f}_{r^*}(r-r^*)}{\left(r-r^*\right)\tilde{f}_{r^*}(r-r^*)}.
 \label{eqn:finalLaplaceTransformAndFPT}
\end{equation}

\section{Inference by extrapolation procedure}

In this section we will describe in detail the inference procedure used in the main text to obtain the unbiased mean FPT values. We will also present alternative procedures we tested and compare their results. All procedures are based on the FPT distribution with Poisson resetting at reset rate $r^*$, obtained from simulations.

The chosen procedure (which we denote as method A) begins with predicting the mean FPT at eight equally spaced rates in the vicinity of $r^*$, $\langle \tau\rangle_{r^*+i\Delta r}$, $i=1,2,...,8$, which was done using Eq. \ref{eqn:finalLaplaceTransformAndFPT} (Eq. 1 of the main text). The results given in the main text use a spacing $\Delta r=0.4r^*$ between adjacent points. Next, the first four derivatives are evaluated at $r=r^*$ using a forward finite difference method, through
\begin{equation}
  \left(\frac{d^n\langle \tau \rangle_r}{dr^n}\right)_{r=r^*} = \frac{\sum_{i=0}^8 C_{n,i} \langle \tau\rangle_{r^*+i\Delta r}}{\left(\Delta r\right)^n}, \label{eqn:finiteDifference}
\end{equation}
where $n$ is the order of the derivative and $C_{n,i}$ are coefficients given in table \ref{tab:finiteDifference}\cite{fornberg_generation_1988}.
These derivatives are used to obtain an approximate fourth-order Taylor expansion of $\langle \tau \rangle_r$ around $r=r^*$, 
\begin{equation}
  \langle \tau \rangle_r = \langle \tau \rangle_{r^*} + \sum_{n=1}^4 
  \left(\frac{d^n\langle \tau \rangle_r}{dr^n}\right)_{r=r^*} \cdot \frac{(r^*-r)^n}{n!} + \mathcal{O}\left((r^*-r)^5\right).
  \label{eqn:Taylor}
\end{equation}
The estimated unbiased mean FPT $\langle \tau \rangle_0$ is simply the value of this function at $r=0$.

\begin{table}
\caption{Coefficients for the finite difference approximations.}
\label{tab:finiteDifference}
\begin{tabular}{|c|c|c|c|c|c|c|c|c|c|}
\hline
i & 0 & 1 & 2 & 3 & 4 & 5 & 6 & 7 & 8 \\ \hline
$C_{1,i}$ & $-\frac{49}{20}$ & 6 & $-\frac{15}{2}$ & $\frac{20}{3}$ & 
$-\frac{15}{4}$ & $\frac{6}{5}$ & $-\frac{1}{6}$ & 0 & 0 \\ \hline
$C_{2,i}$ & $\frac{469}{90}$ & $-\frac{223}{10}$ & $\frac{879}{2}$ & $-\frac{949}{18}$ & 
41 & $-\frac{201}{10}$ & $\frac{1019}{180}$ & $-\frac{7}{10}$ & 0 \\ \hline 
$C_{3,i}$ & $-\frac{801}{80}$ & $\frac{349}{6}$ & $-\frac{18353}{120}$ & $\frac{2391}{10}$ & 
$-\frac{1457}{6}$ & $\frac{4891}{30}$ & $-\frac{561}{8}$ & $\frac{527}{30}$ & $-\frac{469}{240}$ \\ \hline 
$C_{4,i}$ & $\frac{1069}{80}$ & $-\frac{1316}{15}$ & $\frac{15289}{60}$ & $-\frac{2144}{5}$ & 
$\frac{10993}{24}$ & $-\frac{4772}{15}$ & $\frac{2803}{20}$ & $-\frac{536}{15}$ & $\frac{967}{240}$ \\ \hline 
\end{tabular}
\end{table}
 	
\begin{figure}
  \includegraphics[width=\linewidth]{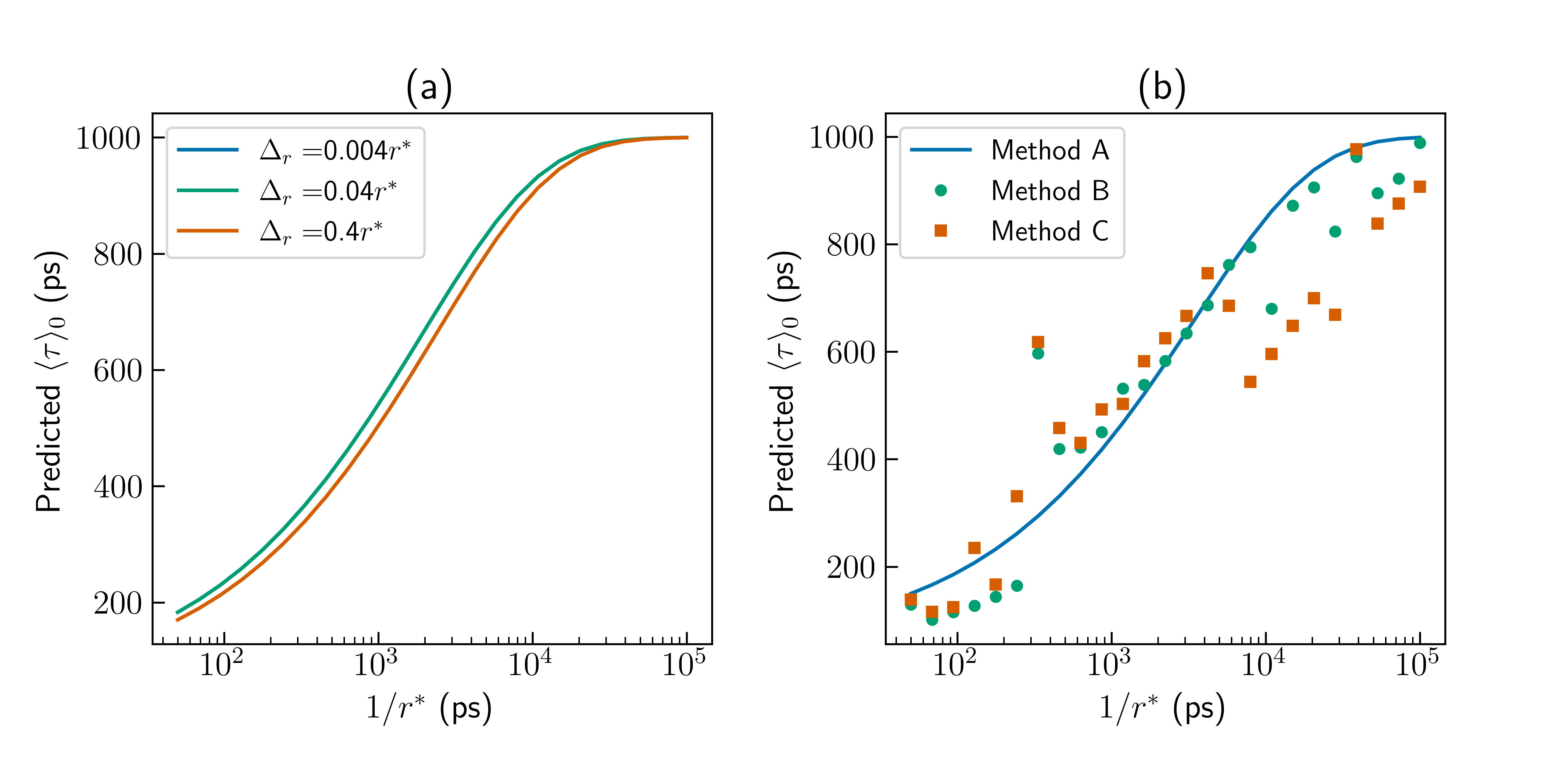}
  \caption{(a) Predictions of $\langle \tau \rangle_{r=0}$ against $1/r^*$ using method A, for different values of $\Delta r$. (b) Predictions of $\langle \tau \rangle_{r=0}$ against $1/r^*$ using method A, B and C. The predictions were made using exact values $\langle \tau\rangle_{r^*+i\Delta r}$ for the inverse Gaussian distribution.}
  \label{fig:inferenceMethods}
\end{figure}

We examined the sensitivity of the extrapolation to the selection of $\Delta r$, and found less sensitivity than in other methods. Fig.\ref{fig:inferenceMethods} (a) shows the predicted $\langle \tau \rangle_0$ against $1/r^*$ for different selected values of $\Delta r$, for the inverse Gaussian distribution with $\langle \tau \rangle_0=1000 \, ps$ using the analytical Laplace transform in Eq.~\ref{eqn:finalLaplaceTransformAndFPT}. It demonstrates that
$\Delta r$ values of different orders of magnitude yield similar predictions.

We also examined fitting directly the mean FPT values at different rates. We will refer to this method as method B. Here, we used a Pad\'e approximant of the form $\langle \tau\rangle_r=\frac{ar^3+br^2+cr +d}{er^2+fr+1}$, which diverges in the limit $r\rightarrow \infty$ as does $\langle \tau\rangle_r$. We fitted the function numerically using Scipy\cite{2020SciPy-NMeth} and substituted $r=0$ to obtain the unbiased mean FPT.

A third method, which we denote as method C, uses a known connection between the mean of a distribution and its Laplace transform, $\langle \tau\rangle=-\left(\frac{d\tilde{f}(s)}{ds}\right)_{s=0}$. We rewrite Eq.~\ref{eqn:originalLaplaceTransformAndFPT} as $\tilde{f}(r) = \frac{1}{1+r\langle \tau \rangle_r}$
and obtain $\tilde{f}(r)$ for several rates $r>r^*$ using $\langle \tau\rangle_{r>r^*}$.
Then, we fit numerically a Pad\'e approximant to these selected values of the Laplace tranform. We use the function  $\tilde{f}(s)=\frac{a+bs^2+cs}{a+ds^3+es^2+fs}$, which fulfills two general properties of Laplace transforms, $\tilde{f}(0)=1$ and $\lim_{s \to \infty}\tilde{f}(s)=0$.
Finally, we evaluate the unbiased mean FPT using the derivative of the fitted 
function at zero, $\langle \tau \rangle_0=\frac{f-c}{a}$.

Fig.~\ref{fig:inferenceMethods} (b) compares between the methods.
Though methods B and C do not require equal spacing, here we used eight equally spaced points as needed for method A. Additional tests did not show any better performance for different spacings or number of points. We used $\Delta r=r^*$ since methods B and C proved to be more sensitive to the selection of $\Delta r$, and performed well for this value. These methods gave similar predictions to those of method A for most values of $r^*$, but deviated for others. Since the predictions of method A improved systematically as $r^*\to 0$, as opposed to the predictions of methods B and C, we choose to present this method in the main text.

\section{Sensitivity to initial conditions and FPT definition}

As discussed in the main text, we checked the sensitivity of the method to the definition of the FPT and to the distribution of initial spatial positions.
The results presented in the main text are for FPT defined as the first crossing of a fixed value $x_1$ close to the minimum of the target basin. We performed additional simulations with different values of $x_1$, and obtained the FPT distributions for these values.
The mean and coefficient of variation (COV) of the unbiased simulations in each case are given in table \ref{tab:passageValues}, along with the maximum speedups we obtained. The value of $x_1$ selected for the results presented in the main text is marked in bold. For all model systems, $x_1=0$ is the peak of the barrier.

\begin{table}
\caption{Mean and COV of the FPT distributions with no resetting for the three models in the main text, with different values defining the first-passage threshold between states. Also included are the maximum speedups gained for both Poisson and sharp resetting.}
\label{tab:passageValues}
\begin{tabular}{|c|c|c|c|c|c|}
\hline
\multirow{2}{1cm}{\textbf{Model}} & 
\multirow{2}{2cm}{\textbf{Passage value ($\AA$)}} &
\multirow{2}{2cm}{\textbf{Mean FPT (ps)}} &
\multirow{2}{1cm}{\textbf{COV}} &
\multicolumn{2}{c|}{\textbf{Speedup}} \\
\cline{5-6}
& & & &\textbf{Poisson} & \textbf{Sharp}\\
\hline
One dimensional double-well
 & 0 &  850 &  3.58 & 20.2 & 18.7 \\ 
 & -1 &  1050 &  3.28 & 16.0 & 14.5 \\ 
 & -2 &  1175 &  3.07 & 13.9 & 12.7 \\ 
 & \textbf{-3} &  1325 &  2.92 &  10.5 & 12.1 \\ 
 & -4 &  1475 &  2.82 & 10.2 & 9.3 \\ \hline
 
 Gimondi et al. &  0 &   875 &  1.54 & 9.1 & 9.9 \\
 & -1/3 &  1075 &  1.44 & 8.9 & 10.1 \\
 & -2/3 &  1125 &  1.44 & 8.2 & 9.4 \\ 
 & \textbf{-1} &  1125 &  1.44 &  8.0 & 9.0 \\ 
 & -4/3 &  1225 &  1.41 & 6.0 & 6.9 \\ \hline
 
 Modified Wolfe-Quapp &  0 &   1050 &  1.44 & 2.0 & 2.2 \\
 & 1/3 &  1100 &  1.43 & 1.9 & 2.0 \\
 & 2/3 &  1125 &  1.43 & 1.9 & 1.9 \\ 
 & \textbf{1} &  1125 &  1.43 &  1.9 & 1.9 \\ 
 & 4/3 &  1125 &  1.41 & 1.9 & 1.9 \\ \hline
 
\end{tabular}
\end{table}

The results are similar for different choices of $x_1$. The change in the mean and COV is less than $50\%$ for the one dimensional potential and less than $10\%$ for the modified Wolfe-Quapp potential.
The obtained speedups are almost identical for all cases of the modified Wolfe-Quapp potential and very similar for the case of the potential of Gimondi et. al. For the one dimensional double-well, the speedup was doubled when defining the passage value as the top of the barrier. It should be noted that we only simulated transitions at a single restart rate, that was expected to give the optimal speedup with Poisson resetting according to the unbiased distribution and Eq.~\ref{eqn:finalLaplaceTransformAndFPT}. It isn't necessarily the optimal rate for sharp resetting, and greater speedups should be expected for sharp resetting when optimizing the restart rate.

The results of the main text used fixed spatial initial conditions. This is equivalent to sampling the positions initially, and after each reset, from a delta function distribution. We examined the influence of the choice of distribution by simulating trajectories with positions sampled from the
Boltzmann distribution at the beginning of the simulations and after each reset event. The results are given in table \ref{tab:boltzmanSample}.

\begin{table}
\caption{Mean and COV of the FPT distributions with no resetting for the three models in the main text, with initial positions sampled from the Boltzmann distribution. Also included are the expected maximum speedups for both Poisson and sharp resetting.}
\label{tab:boltzmanSample}
\begin{tabular}{|c|c|c|c|c|}
\hline
\multirow{2}{1cm}{\textbf{Model}} & 
\multirow{2}{2cm}{\textbf{Mean FPT (ps)}} &
\multirow{2}{1cm}{\textbf{COV}} &
\multicolumn{2}{c|}{\textbf{Speedup}} \\
\cline{4-5}
& & &\textbf{Poisson} & \textbf{Sharp}\\
\hline
One dimensional double-well & 6525 &  1.24 & 3.5 & 3.7 \\
Gimondi et al. & 1350 &  1.27 & 3.3 & 3.3 \\
Modified Wolfe-Quapp & 1750 &  1.12 & 1.5 & 1.5 \\
\hline
\end{tabular}
\end{table}

The mean FPT is greater than the one achieved with fixed initial positions, and the COV is lower. As expected, the speedups are lower as well, because there is a significant probability to initiate the simulations very far from the barrier. Nevertheless, The COV remained greater than one and speedup was obtained using stochastic resetting in all model systems. This verifies that the acceleration gained by SR is not dependent on using a single specific initial condition. 





\bibliography{si}